\documentclass[aps,prd,reprint,superscriptaddress,floatfix]{revtex4-1}

\usepackage[colorlinks,allcolors=blue]{hyperref}

\usepackage{graphicx}
\usepackage{rotating}
\usepackage{multirow}
\usepackage{amsmath}
\usepackage{amsfonts}
\usepackage{subcaption}
\usepackage{comment}

\newcommand*\diff{\mathop{}\!\mathrm{d}}

\usepackage{xspace}

\begin{document}

\title{Deep-learning-based low-energy trigger algorithms\\for the Hyper-Kamiokande experiment}

\author{Katharina Lachner}
\affiliation{ETH Z\"urich}
\affiliation{University of Warwick}
\author{Sa\'ul Alonso-Monsalve}
\affiliation{ETH Z\"urich}
\author{Benjamin Richards}
\affiliation{University of Warwick}
\author{Davide Sgalaberna}
\affiliation{ETH Z\"urich}

\begin{abstract}
    Modern machine learning techniques have become increasingly important in particle
    physics because of their powerful pattern-recognition capabilities, including in
    real-time data acquisition where stringent runtime constraints apply.
    This paper details the
    performance of deep-learning-based trigger algorithms for a large water
    Cherenkov detector such as Hyper-Kamiokande aimed at low-energy neutrino events
    (below 7\,MeV). The performance of custom neural-network supervised classifiers is shown alongside two anomaly-detection approaches trained solely on detector noise: a pure autoencoder and an energy-based model based on Manifold Projection--Diffusion Recovery (MPDR). The supervised model shows signal identification efficiencies of
    76.7\% for single electrons of 3\,MeV kinetic energy, significantly
    exceeding signal efficiencies obtained from a traditional hit-count-based trigger
    of 26.4\%, as does the MPDR approach with 31.8\%. Runtime
    evaluations on GPU yield per-window inference
    latencies well below the millisecond scale, indicating that real-time
    operation is feasible.
\end{abstract}

\maketitle

\section{Introduction}

The Hyper-Kamiokande experiment comprises a vast physics programme with a notable contribution from neutrino studies in the few-MeV range. The far
detector, a Cherenkov detector located 0.65\,km (1.75\,km water equivalent)
underneath Mount Nijugo near Kamioka in Gifu, Japan, will hold a fiducial mass of
187\,kton of ultra pure water. Equipped with state-of-the-art Box-and-Line Photomultiplier Tubes~(PMTs), it will provide a low-background detection environment, permitting low energy neutrino measurements at unprecedented
precision. However, the success of physics studies in this energy range not
only relies on hardware developments: the low end of the detectable neutrino
energy spectrum also depends heavily on the performance of Data Acquisition~(DAQ) and triggering techniques that target the identification of such signals.
Development and optimisation of dedicated low-energy trigger algorithms is thus
crucial to ensure Hyper-Kamiokande can fully exploit its physics potential~\cite{abe2018hyper}.

The Hyper-Kamiokande DAQ system incorporates a fully software-based
approach, with an on-site data centre to take care of real time data processing.
This allows for versatile and continued development of optimised trigger
algorithms. Modern machine learning methods are well suited
for the task of identifying signals from low-energy neutrino interactions in
this region where background dominates the signal, as early studies of machine-learning-based triggers indicated~\cite{dealtry2018tmex}. 
In this paper, different machine-learning-based implementations of
trigger algorithms are described. These aim at optimising signal identification at
electron-equivalent kinetic energies below 7\,MeV.

Custom low-energy deep-learning trigger algorithms for the Hyper-Kamiokande far detector are presented, including supervised transformer-encoder classifiers as well as anomaly-detection methods.
In addition to signal identification efficiencies for single
electron signals and capabilities towards gamma identification used by
neutron tagging algorithms in water Cherenkov detectors, we show runtime estimates for these algorithms on a CPU and
GPU, and discuss the feasibility of either of the algorithms as part of an
online triggering suite such as is being developed by the Hyper-Kamiokande
experiment.

\section{Motivation}

Like its predecessor Super-Kamiokande~(Super-K)~\cite{fukuda2003super}, the Hyper-Kamiokande
experiment~(Hyper-K) is designed as a multi-purpose experiment, targeting
measurements across a wide range of neutrino energies: from rare cosmogenic
neutrinos at high energies in the TeV range, $CP$-violation studies with
accelerator and atmospheric neutrinos in the GeV range to studies of solar, supernova, and reactor neutrinos at typical energies close to the detection threshold of a few
MeV~\cite{abe2018hyper}. The latter are of particular interest to help resolve observed tensions
in $\Delta m^2_{21}$ measurements from reactor and solar neutrino studies, which
mainly stem from day-night-asymmetry analyses at Super-K. A repetition of such a
measurement at higher precision is required to either help resolve
these tensions between the solar $\nu_e$ and reactor $\bar{\nu}_e$ measurements, or
otherwise confirm potential hints towards new physics~\cite{gando2013reactor,abe2024solar,esteban2024nufit}. The majority of solar neutrinos are emitted with energies below 10\,MeV, motivating a focus on increasing efficiencies at lowest detectable range~\cite{bellerive2004review}.
The improved time resolution and larger volume of Hyper-K will enable enhanced background
rejection techniques, providing a strong basis for such measurements.
Moreover,
studies of sterile neutrinos and other exotic models would benefit from a first
measurement of the MSW-induced upturn in the solar neutrino survival probability. This upturn reflects the transition of the oscillation probability from matter-dominated behaviour at higher energies to vacuum oscillations at lower energies, motivating a
dedicated focus on the low-energy end of the measurable neutrino
spectrum~\cite{abe2018hyper}.

Studies of neutrinos created in supernovae represent another aspect of the
low-energy physics program in Hyper-K. Particle dynamics in the core of
collapsing stars are currently not fully understood, and details of the
mechanism triggering supernovae are still an active area of research~\cite{bethe1985revival}.  
With SN1987A being the only instance of a core-collapse supernova from which the corresponding neutrino flux was observed,
the measured rate of 24 candidate
events in the Kamiokande, IMB, and 
Baksan experiments only provides limited statistics for data analysis~\cite{hirata1987observation,
bionta1987observation, alexeyev1988detection}.  While Hyper-K is preparing for a
potential observation of a nearby core-collapse supernova, 
an event expected to occur only once or twice every century~\cite{diehl2006radioactive}, a more
prevalent handle on neutrinos created by supernovae is the Diffuse Supernova
Neutrino Background~(DSNB)~\cite{totani1995spectrum,malaney1997evolution,lunardini2026diffuse}. Such an integrated
flux of neutrinos created in supernovae throughout the universe is yet to be
confirmed by experiment~\cite{abe2021diffuse,harada2023search}. Predicted flux and spectra vary between theoretical models, but generally peak at around 5\,MeV~\cite{vagins2012detection}. The $\bar{\nu}_e$ component of the DSNB flux provides a good handle on these events, as interactions via Inverse Beta Decay (IBD) can occur on free protons. Neutrons created in such interactions can prove useful in the identification of anti-neutrino signals, as has been demonstrated by the Super-K experiment~\cite{watanabe2009first}. Hyper-K could make a first measurement of the DSNB possible
if backgrounds in the few MeV range are well understood and if a sufficient
neutron tagging efficiency can be achieved. A measurement of the DSNB would not
only provide inputs to supernova models, but could also give access to stellar
history and physics beyond the Standard Model if observed spectrum shapes differ
from predictions.

It is crucial for Hyper-K to make the best use of its potential regarding
measurements in the low end of the neutrino energy spectrum, and to push the detection threshold below 7\,MeV.  This
motivates the development of dedicated pattern recognition tools for the task of
identifying Cherenkov signals  over the dominating detector noise.
Such developments have to begin already at the level of data acquisition to
ensure detector readout containing the relevant energy region will be
saved to disk for more detailed offline analyses of low-energetic
extraterrestrial neutrinos~\cite{abe2018hyper}.

Across particle physics, machine-learning-based selection has increasingly moved
``upstream'' into real-time pipelines, driven by the same pressure Hyper-K faces
at low energies: a region of interest dominated by noise,
limited bandwidth, and the need to maximise the efficiency for rare
signatures. Community efforts on fast/real-time ML, including GPU- and
FPGA-oriented deployments, show that modern architectures can be engineered to
satisfy strict throughput/latency constraints while retaining much richer
decision functions than traditional hand-tuned
triggers~\cite{Astrand_2026,deiana2022fastml,aaij2020allen,duarte2018fpga}. In
particular, fully software-based triggers operating at collider-scale rates on
heterogeneous CPU/GPU farms demonstrate that high-capacity models can be
integrated into production DAQ environments when the inference path is designed
with performance in mind~\cite{aaij2020allen}.

For the low-energy program in a large water-Cherenkov detector, the
discriminating information is predominantly encoded in sparse spatiotemporal
PMT-hit correlations (ring-like topology, timing coherence, and local
charge/time structure). Deep learning is well suited to this regime because it
can learn multi-dimensional correlations directly from low-level detector
readout, while naturally handling variable hit multiplicity. In particular,
set/point-cloud formulations avoid lossy voxelisation and preserve detector
granularity; self-attention layers provide a principled way to aggregate
information across all hits and learn long-range correlations, and
transformer-based point-cloud models have demonstrated strong performance in
both computer vision and particle-physics
applications~\cite{zhao2021pointtransformer,qu2020particlenet}. Finally, because
trigger decisions must remain robust to potential signal mismodelling, it is
well-motivated to complement supervised discrimination with noise-only
(anomaly-detection) strategies, a direction that has received growing attention
in particle physics and has even reached real-time demonstrators in modern trigger
systems~\cite{belis2024anomalyreview,gandrakota2024cmsl1}.

\section{Input preparation} 
\label{sec:input}

This section details the simulation of low-energetic neutrino events in the
Hyper-K far detector. The geometry of the detector is described first, followed
by specifications of the training and testing samples used in this study.

\subsection{Detector simulation}

\begin{figure*}[htb]
    \begin{subfigure}[t]{0.31\textwidth}
        \includegraphics[height=.75\textwidth,clip,trim={30mm 0 15mm 0}]{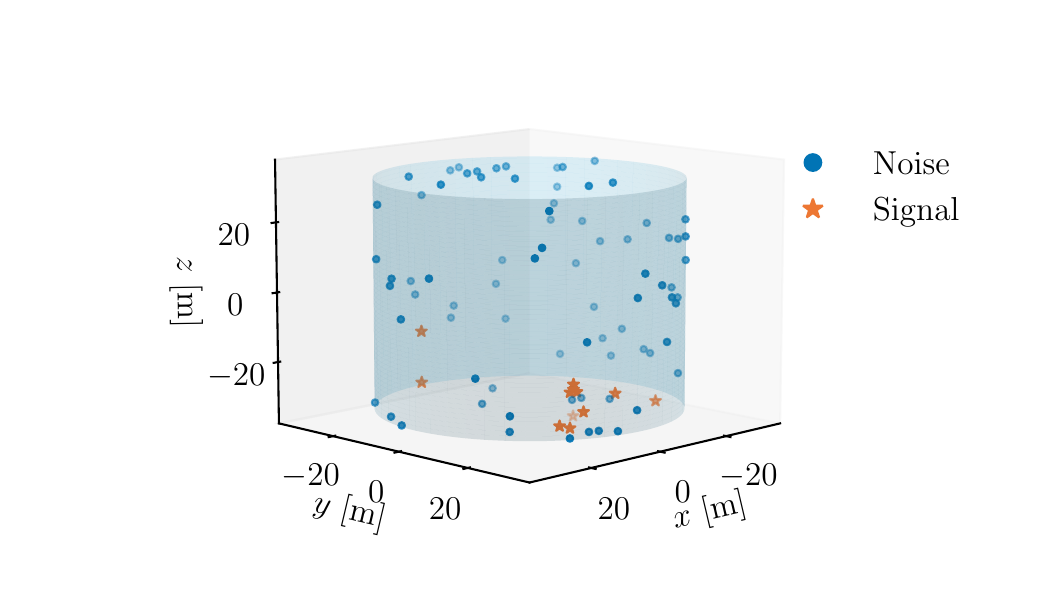}
        \vspace{-2em}
        \caption{ }
    \end{subfigure}
    \begin{subfigure}[t]{0.31\textwidth}
        \includegraphics[height=.75\textwidth,clip,trim={40mm 0 10mm 0}]{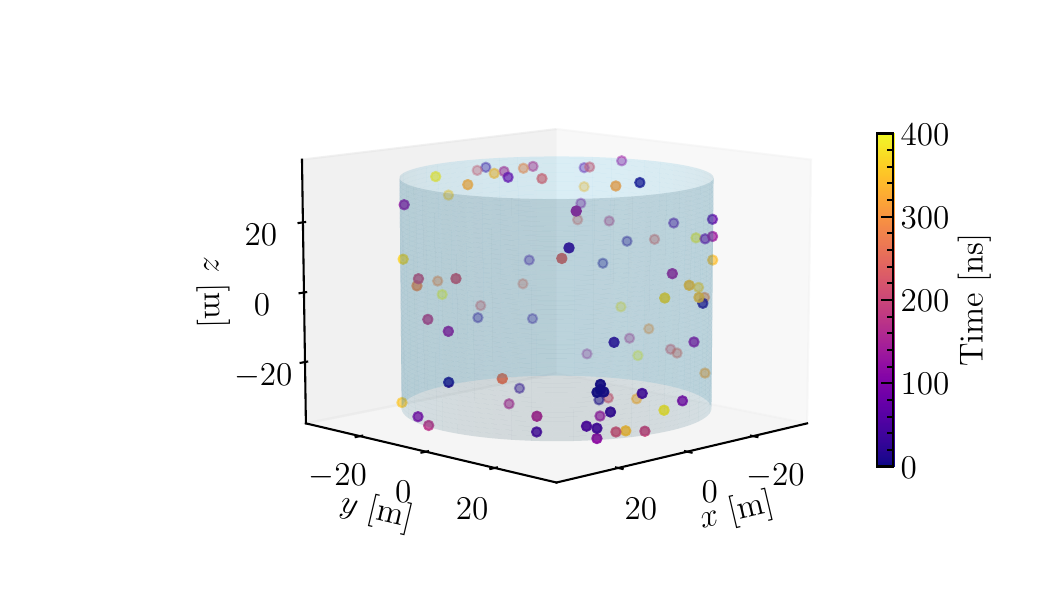}
        \vspace{-2em}
        \caption{ }
        \hspace{0.2cm}
    \end{subfigure}
    \begin{subfigure}[t]{0.31\textwidth}
        \includegraphics[height=.75\textwidth,clip,trim={40mm 0 15mm 0}]{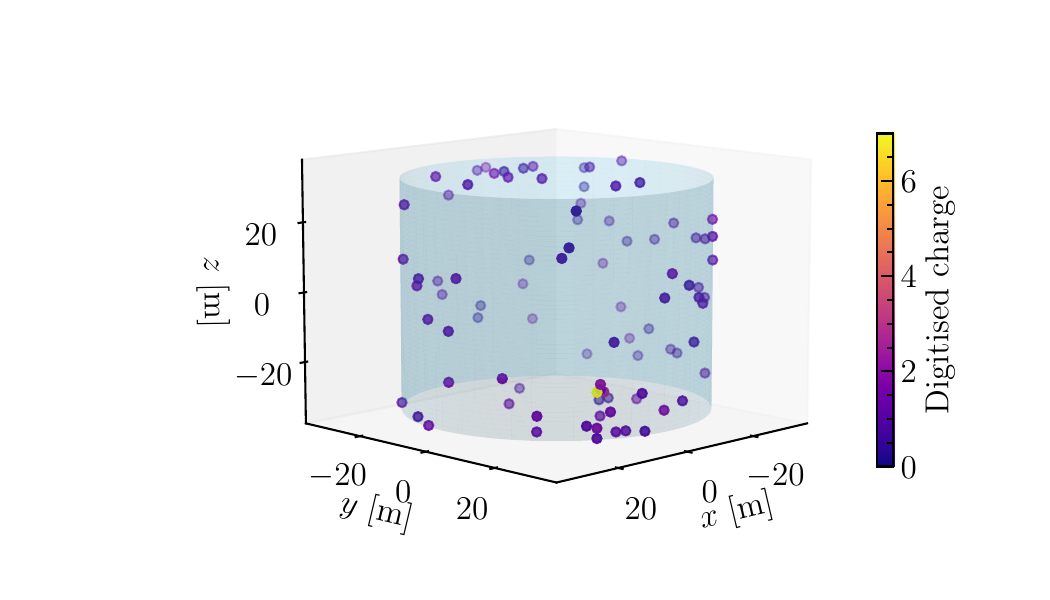}
        \vspace{-2em}
        \caption{ }
    \end{subfigure}
    \vspace{-1em}
    \caption{\label{fig:evt_disp}
        Event display for an example $\nu_e$ interaction creating an electron of
        3\,MeV kinetic energy, colour coded by signal vs. noise hits (a), hit
        time (b), and readout charge (c).
    }
\end{figure*}

The Geant4-based~\cite{agostinelli2003geant4} open source simulation framework WCSim~\cite{dealtry2016wcsim}
is utilised to simulate the response of the Hyper-Kamiokande far detector. A
cylinder of 32.4\,m radius and 65.9\,m height comprises the Inner Detector (ID),
and 19\,746 Box-and-Line PMTs of 20\,inch diameter cover the walls
and endcaps at a photo-coverage of 20\%. Hyper-Kamiokande will also be equipped
with additional so-called Multi-PMT modules~\cite{derosa020multi}, but only Box-and-Line PMTs are
considered in this study.  Information from PMTs in the Outer Detector (OD) is
disregarded as well for simplicity.  PMT noise rates are conservatively
estimated at a rate of 10\,kHz per PMT.  All results in this paper use WCSim
version~1.12.24 based on Geant4 version 10.3.3, using the \texttt{FTFP\_BERT}
physics list.

The WCSim framework provides a
realistic simulation of optics in the detector, including scattering and absorption
phenomena as well as reflections.  It also simulates wavelength-dependent
collection and quantum efficiencies in the PMTs, set to a maximal value
of ${QE=0.211}$ at 400\,nm. 
The time resolution is expected to be limited by the PMT response, which is simulated as a charge-dependent randomised time offset (jitter) with a dependence on hit charge $Q$ and the Transit Time Spread~(TTS), the latter being assumed at 1\,ns. A Gaussian distribution of the following width is considered:
\begin{equation} 
    \sigma_t(Q) = \mathrm{TTS}\cdot\left( 0.33 + \sqrt{\frac{10}{Q}} \right)
\end{equation}
A cut-off for the minimum time resolution is applied at 0.58\,ns. Timing resolutions of digitiser and electronics are higher, and all values are kept at the WCSim default for 20\,inch PMTs.

\subsection{\label{sec:evt_characteristics}Event characteristics}

Single electrons are simulated at a random time within a 400\,ns window.
This duration is chosen to be slightly larger than the light propagation time across the detector diagonal, ensuring full event containment if events occur at the beginning of the window. The particles are placed at random positions and assigned random directions in the
volume of the ID and up to 1\,m outside of the wall separating ID and OD. Vertices outside the ID volume are included to account for leakage effects, as such events can be indistinguishable from interactions originating within the ID in real data. The
training sample is made up of 1 million events of single electrons with kinetic energies uniformly distributed between 0 and 7 MeV, split into training and validation sets using an 80/20
ratio. The trigger decision is then evaluated on this single 400\,ns time window. It should be noted that during training we do not consider consecutive time windows, and as
a result not necessarily all detectable hits fall within the saved time frame of
0-400\,ns, which emulates realistic conditions.  We thus additionally require at
least one hit caused by Cherenkov light emitted by the primary electron to
define an event as a true signal, and also apply per-hit tags to separate signal
and noise hits at truth. The testing samples for performance evaluation are
chosen as mono-energetic electron events, again at random positions and
directions in the same volume as the training sample, however with interaction
times set to the beginning of a 1\,$\mu$s time window to ensure the full event is covered.  Such samples are created
for 0.5\,MeV to 7\,MeV in steps of 0.5\,MeV, with 10\,000 events per energy
slice.  Trained models are then applied to evaluate the trigger decision on
three slices: 0-400\,ns, 300-700\,ns, and 600-1000\,ns.  

For a study towards neutron tagging capabilities, we also evaluate the signal identification performance for single gammas of 2.2\,MeV energy. The capture of the neutron on free hydrogen leads to the emission of such a mono-energetic photon which happens at an average of 205\,$\mu$s after the primary lepton
signal. Such photons can be detected via Compton
scattering, producing recoil electrons at or below MeV energies. For the purpose of this paper, we do not consider temporal correlation with the prompt lepton, and for this testing sample only the gamma is simulated. The sample contains 10\,000 events, and like the electron sample, the particles are placed at random positions up to 1\,m outside the ID volume, and are assigned
directions. Models trained exclusively on single electrons are then applied on this sample.  However, it should be
noted that a dedicated neutron tagging algorithm could
likely outperform results shown here by considering the temporal correlation with the prompt lepton signal.

\begin{figure}[htb]
    \centering
    \includegraphics[height=.30\textwidth]{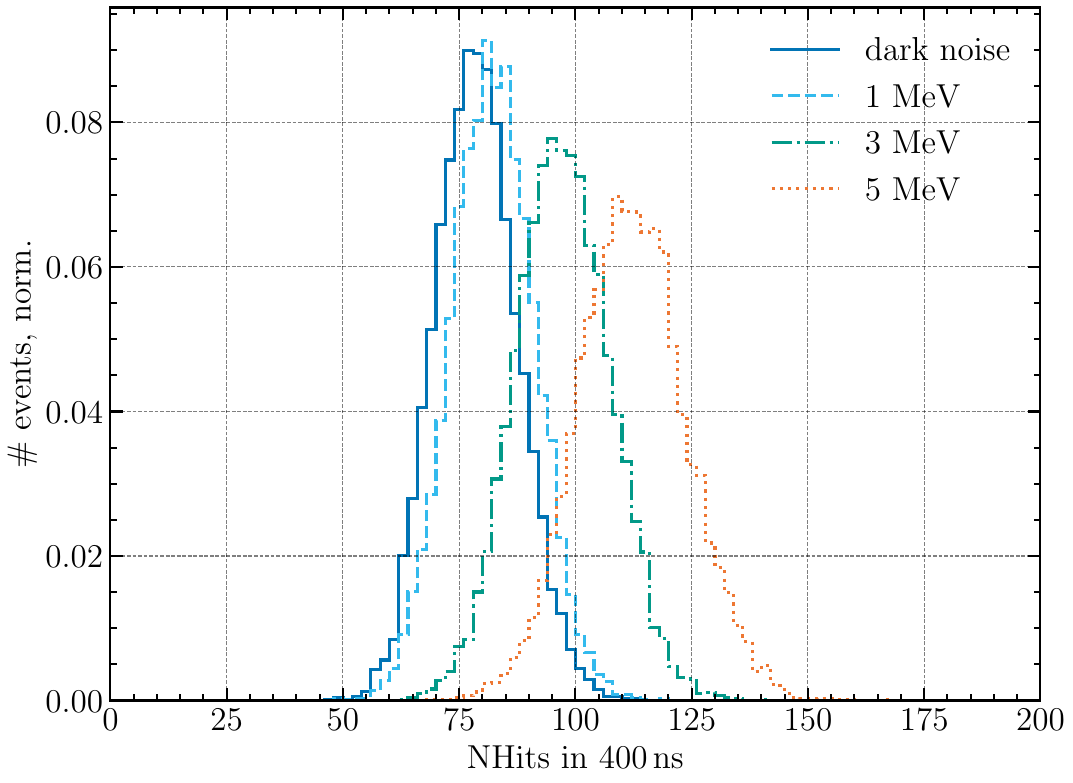}
    \caption{\label{fig:nhits}Distribution of the number of hits in a 400\,ns
        slice for pure detector noise at 10\,kHz rate, as well as interactions at the beginning of
        the slice of mono-energetic electrons at 1\,MeV, 3\,MeV, and 5\,MeV on
        top of the detector noise.
    }
\end{figure}

\begin{figure*}[htb]
    \begin{subfigure}[t]{0.4\textwidth}
        \includegraphics[height=.6\textwidth]{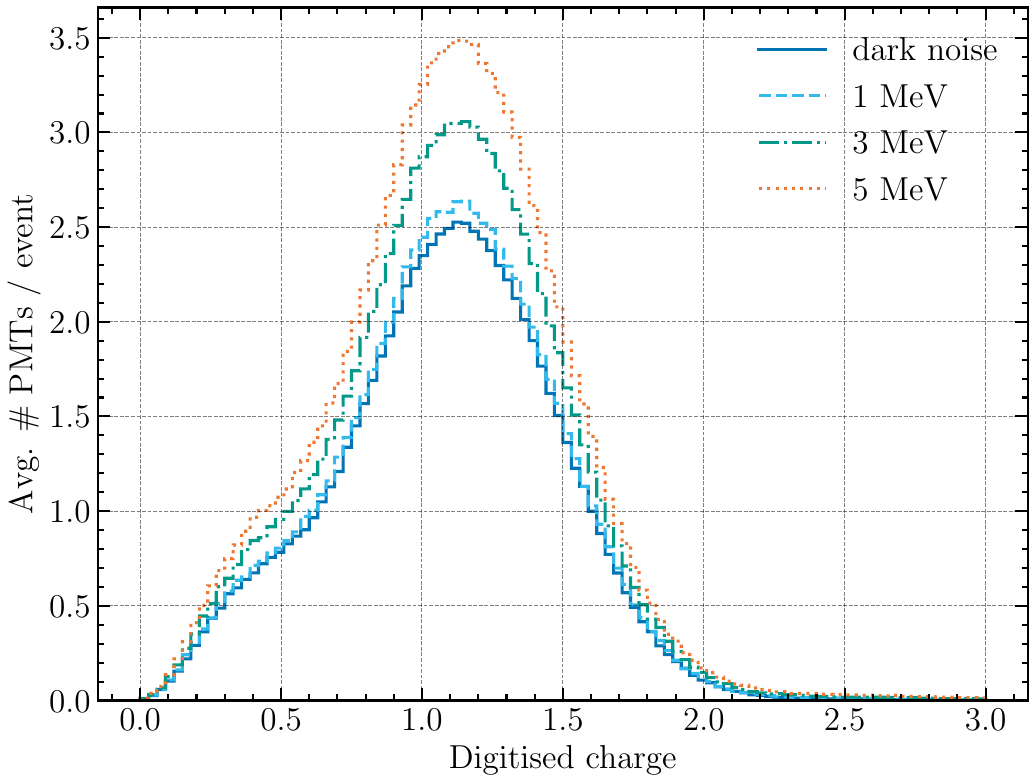}
        \caption{ }
    \end{subfigure}
    \hspace{1em}
    \begin{subfigure}[t]{0.4\textwidth}
        \includegraphics[height=.6\textwidth]{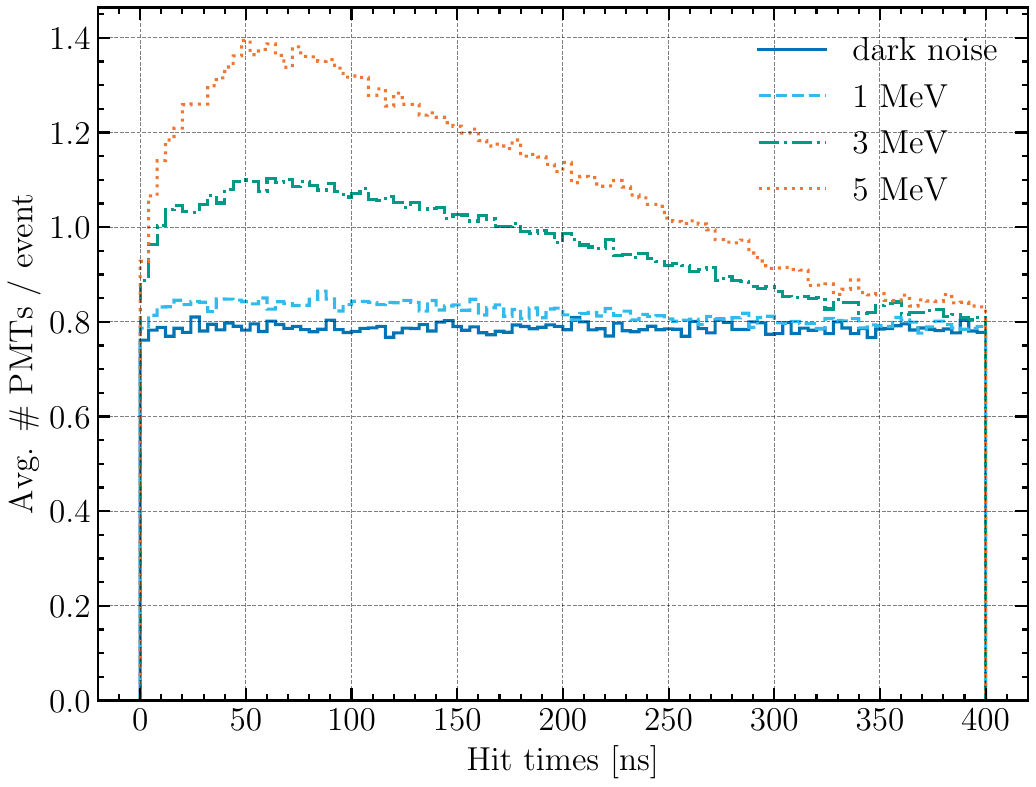}
        \caption{ }
    \end{subfigure}
    \caption{\label{fig:training_sample_distrs}
        Distributions of (a) hit charges and (b) hit times in the first 400\,ns in pure noise as well as immediately post-interaction for mono-energetic electrons at 1\,MeV, 3\,MeV, and 5\,MeV on top of the noise.
    }
\end{figure*}

Typical high-energy triggers in Cherenkov detectors operate based on a cut on the
number of hits, an algorithm which in the following will be referred to as the
\texttt{NHits trigger}. Figure~\ref{fig:nhits} shows the distribution of active
PMTs in a 400\,ns pure noise sample compared to single electrons at 1\,MeV,
3\,MeV, and 5\,MeV on top of simulated noise. For low-energy events, a
significant overlap in the number of hits from signal and noise samples is
evident.  Constraints on the false trigger rate have to be placed to ensure
real-time data processing is possible, which results in low efficiencies using
such an NHits trigger for low-energetic events. Hence, more sophisticated
approaches are required to distinguish between low-energy signal and noise.  An
example event for an electron of 3\,MeV kinetic energy is presented in
Fig.~\ref{fig:evt_disp}a, with colour code highlighting PMTs that measured
Cherenkov light (``signal''), and PMTs active due to detector noise. Figures~\ref{fig:evt_disp}b and~\ref{fig:evt_disp}c show the same event with colour code indicating time and charge,
respectively.  
In particular, correlations in signal hit times can be observed,
while signal hit charges tend to remain at the level of noise hits for faint
Cherenkov signatures from few-MeV leptons. Spatial and temporal
correlations, and potentially higher charge readout thus represent the patterns
that machine-learning models presented in this paper should recognise. 
Characteristic charge and time distributions are shown in Fig.~\ref{fig:training_sample_distrs} for pure noise as well as mono-energetic electrons at 1\,MeV, 3\,MeV, and 5\,MeV. An increase in the number of hits at around 50-100\,ns post-interaction can be observed, while no significant shift towards higher averages of digitised charges is visible with increasing electron energies in this range.

\section{Deep-learning approach}
\label{sec:method}

We address the challenge of triggering low-energy events in the Hyper-Kamiokande
detector through two complementary approaches: supervised transformer-based
classifiers for direct signal-noise discrimination, and unsupervised methods for anomaly detection. Both methods operate on sparse point cloud representations of Water-Cherenkov photomultiplier tube (PMT) hit patterns, leveraging the geometric and temporal structure of the detector data. All models presented in this paper are publicly available at \url{https://github.com/saulam/hk-ml-trigger}.

\subsection{Data representation}

For all trigger algorithms, an input example corresponds to a fixed readout window and is represented as a sparse, variable-length set of PMT hits (a point cloud). Each hit is associated with (i) the PMT position on the detector surface, (ii) the hit time within the readout window, and (iii) the integrated charge. The supervised classifier additionally uses (iv) a discrete PMT-location label that distinguishes top-cap, barrel, and bottom-cap PMTs; this is used in the geometric feature construction and token-type embedding.

Treating events as point clouds preserves the detector's native spatial granularity and avoids dense voxelisation. Variable hit multiplicities are handled by padding and attention masking during batching.

While all models start from the same raw hit information, we use two different geometric parametrisations to match each method's needs.
{For the supervised transformer classifiers, hit locations are encoded in cylindrical coordinates aligned with the Hyper-K detector axes. This makes the barrel symmetry explicit and supports geometry-aware attention. For anomaly detection, we instead use a spherical reparametrisation of detector-surface position: each hit at Cartesian location $\mathbf{x}$ is represented by the full unit direction vector $\mathbf{u}=\mathbf{x}/\|\mathbf{x}\|$.
This choice is made for numerical convenience in the autoencoder and anomaly-detection pipeline (Sec.~\ref{sec:mpdr}). In particular, reconstructed or recovered points can be mapped back to a valid surface representation by a simple normalisation step, avoiding explicit projection onto the piecewise-defined cylindrical detector geometry (barrel and endcaps).} 
Additionally, hit charges are omitted as input features for the anomaly-detection models. We do not interpret the performance difference between the supervised transformer and the anomaly-detection models as arising solely from this coordinate choice, since the methods also differ in supervision, objective, architecture, and input features.

All normalisations, feature definitions, and implementation details for the supervised classifiers are provided in Appendix~\ref{sec:appendix_transformer}. The corresponding details for the autoencoder-based approaches are given in Appendix~\ref{sec:appendix_anomaly}.

\subsection{Supervised classifier}
\label{sec:supervised_classifier}

The supervised approach learns to discriminate signal-like PMT activity from detector noise using labelled simulation. We employ a transformer encoder~\cite{vaswani2017attention} operating directly on the hit point cloud. Self-attention allows the model to combine local hit correlations (e.g., clustered Cherenkov photons in space and time) with global event context, which is essential at energies of a few MeV where the total hit count alone is only weakly discriminative.

To respect the detector geometry, hits are embedded using cylindrical coordinates whose axes are aligned with the Hyper-K detector axes. The azimuthal coordinate is represented by $\sin\phi$ and $\cos\phi$, which makes its periodicity explicit, while additional sinusoidal (Fourier) features are applied to the continuous spatial and temporal coordinates as a multiscale basis expansion. For the time coordinate, these features are used to provide a richer representation over the bounded readout window, rather than to impose any physical periodicity.
In addition, the attention mechanism is augmented with a learnable relative position bias~\cite{shaw2018self} constructed from pairwise differences in cylindrical coordinates and hit times. This provides an explicit inductive bias for translation- and rotation-equivariant pattern recognition in the barrel while retaining sensitivity to endcap topology. Wall and endcap PMTs are treated with separate input projections to account for their different geometric roles.

In this work, we evaluate two independent modelling strategies distinguished by their level of supervision:
\begin{itemize}
    \item \textbf{Event-level supervision:} The model is trained to directly
        predict a single classification label (signal or noise) for the entire event.
    \item \textbf{Hit-level supervision:} The model is supervised at the hit level to identify individual signal hits. For event-level inference in this configuration, the final classification score is defined as the maximum output probability across all constituent hits in the event.
\end{itemize}

The architecture for both approaches consists of a 6-layer encoder with hidden dimension $d_{\text{model}}=64$ and $n_{\text{head}}=8$ attention heads (approximately $3\times10^5$ trainable parameters). Training is performed using binary cross-entropy and data augmentations detailed in Appendix~\ref{sec:appendix_augmentations}, including random rotations about the detector axis and random reflections along each coordinate axis, to improve robustness.

A complete specification of the input normalisation, per-hit feature construction, relative positional attention, ablation feature sets, and training hyperparameters is given in Appendix~\ref{sec:appendix_transformer}.

\subsection{Anomaly detection: autoencoder and MPDR models}
\label{sec:mpdr}

In this approach we treat low-energy triggering as a noise-only learning
problem: we train solely on noise events and then flag test events as anomalous
if they are inconsistent with the learned noise distribution. The Manifold Projection-Diffusion Recovery (MPDR) framework~\cite{yoon2023energybasedmodelsanomalydetection} provides a practical way to implement this idea using an energy-based model,
{i.e.\ a model that learns a scalar scoring function $E_\theta(X)$ over inputs $X$. Here, ``energy'' is purely a technical construct and has no correspondence to the physical energy}; it is a learned quantity for which lower values are assigned to events that are more compatible with the learned data distribution, and higher values to less compatible events.
MPDR is proposed as an extension of recovery-likelihood training. 
It supports both a scalar-energy variant (MPDR-S) and a reconstruction-error energy variant (MPDR-R). The latter follows the reconstruction-energy idea introduced in the Normalized Autoencoder (NAE) work~\cite{yoon2021autoencoding} and differs mainly in the training procedure~\cite{yoon2023energybasedmodelsanomalydetection}. 
In this work we use only MPDR-S and, for brevity, refer to it simply as ``MPDR'' throughout.
{A comparison between MPDR-S and MPDR-R is left for future work.}

The core idea is to learn the typical noise distribution well enough that events compatible with the noise distribution receive low energy, while events containing additional structure receive higher energy and hence larger anomaly scores. This is attractive in our setting not only because it reduces dependence on signal modelling, but also because the noise distribution can in principle be characterised directly from data in detector conditions where no true Cherenkov signal is expected.
MPDR combines two ingredients: first, a manifold autoencoder~$(\mathcal{E},\mathcal{D})$
is trained on noise events to provide an approximate noise manifold. The encoder maps an input event X to a latent representation $z = E(X)$, and the decoder reconstructs the event from $z$. In our implementation, we enforce a unit-norm constraint on the
latent code, $z$, as $\mathbf{z}\leftarrow \mathbf{z}/\|\mathbf{z}\|$, which improves numerical stability
and simplifies the latent perturbation step. Second, an energy function is trained to
distinguish real noise events from ``hard'' synthetic events obtained by perturbing noise
examples in latent space and then partially recovering them into realistic detector hit patterns.
Events containing signal-like structure are expected to fall outside the region of low energy and hence acquire larger scores.

In our implementation, each PMT hit is represented by its position on the detector surface and its time. For the spatial part we use a unit-sphere parametrisation: a hit position $\mathbf{x}\in\mathbb{R}^3$ is mapped to
$\mathbf{u}=\mathbf{x}/\|\mathbf{x}\|\in\mathbb{S}^2$,
and we concatenate the (normalised) hit time $t$, giving per-hit features $(\mathbf{u},t)$.
This choice is primarily practical rather than physics-motivated. During training and, crucially, during MPDR sampling, the model must repeatedly update hit coordinates using gradients while keeping them on a valid detector-surface representation. Using unit-norm spatial coordinates allows this constraint to be enforced by a simple normalisation step, which is smooth almost everywhere and leads to stable gradients. By contrast, enforcing the cylindrical detector geometry directly would require explicit projection of decoder outputs onto a piecewise-defined surface (barrel versus endcaps), together with case distinctions near their boundary. We therefore use the unit-vector representation as a numerically robust surface parameterisation for the autoencoder and MPDR pipeline.

\subsubsection{Phase I: PMT-noise manifold learning with an autoencoder}

In the first phase we train a transformer-based autoencoder $(\mathcal{E},\mathcal{D})$ using pure noise events only. The encoder $\mathcal{E}$ maps a variable-length hit set $X$ to a fixed-dimensional latent code $\mathbf{z}$, and the decoder $\mathcal{D}$ reconstructs the event from $\mathbf{z}$. We constrain the latent code to the unit hypersphere, which stabilises training and provides a convenient space in which to perturb events when constructing MPDR negatives.

To handle the fact that events contain a variable number of hits, the decoder produces a fixed number of candidate hits together with an ``existence'' probability for each candidate. These candidate hits are not additional physical hits inserted into the training sample. They form a fixed-size decoder output used to represent variable-cardinality events: each candidate is assigned an existence probability. The reconstruction objective matches the spatial and temporal distributions while encouraging the correct event multiplicity. The full loss, including the multi-scale point-cloud matching and the treatment of existence probabilities, is given in Appendix~\ref{sec:appendix_anomaly}.

\subsubsection{Phase II: learning an energy function with MPDR negatives}

After the autoencoder is trained, its parameters are frozen and we train an
energy function $E_\theta(X)$ (MPDR-S in
Ref.~\cite{yoon2023energybasedmodelsanomalydetection}). The purpose of
$E_\theta$ is to assign low energy to genuine noise events and higher energy to
events that look plausible but are not drawn from the noise distribution.

Since we do not use labelled signal, the training procedure constructs
\emph{negative examples}, i.e.\ synthetic events that are not genuine noise but
are designed to resemble it closely. Conceptually, MPDR proceeds as follows
(implementation details in Appendix~\ref{sec:appendix_anomaly}). A noise event
is encoded to $\mathbf{z}$, perturbed by adding noise in latent space, and
decoded back to hit space to produce an initial synthetic event. This event is
then refined through a small number of stochastic gradient-based updates,
corresponding to a short Langevin-style Markov Chain Monte Carlo (MCMC) recovery procedure guided by the current energy model, with an additional term that keeps the event from
drifting too far from the original noise example. The resulting negative examples are
deliberately challenging: they resemble PMT noise at the level of basic hit
patterns, but are systematically displaced from the region occupied by genuine
noise events in the learned representation. Training then encourages
$E_\theta$ to assign lower energy to true noise events than to these synthetic
near-noise examples.

At inference time we use
$s(X)=E_\theta(X)$
as the anomaly score so that larger scores correspond to more anomalous events: events that contain additional coherent structure not
captured by the noise model are expected to yield larger scores and can be selected by thresholding $s(X)$.

\section{Results}

As described in Sec.~\ref{sec:method}, models are trained on inputs corresponding to windows of duration $400~\mathrm{ns}$, while performance is evaluated on events defined over $1~\mu\mathrm{s}$ windows.
{We therefore evaluate each test event on three partially overlapping $400~\mathrm{ns}$ sub-windows with start times $0~\mathrm{ns}$, $300~\mathrm{ns}$, and $600~\mathrm{ns}$. This reduces sensitivity to the exact time of the interaction within the $1~\mu\mathrm{s}$ readout window while keeping the inference conditions matched to training. This is particularly relevant because the signal can appear anywhere within the saved readout window.}
We define the event-level score as
\begin{equation}
s \equiv \max_{k\in\{1,2,3\}} s^{(k)}.
\end{equation}
where $s^{(k)}$ is the score for sub-window $k$.
This choice reflects a trigger that fires if any sub-window within the event is sufficiently signal-like (or anomalous), and ensures consistency between training and evaluation windowing.

As a first performance evaluation, 
we quantify discrimination using the area under the receiver operating characteristic curve (AUROC),
\begin{equation}
\mathrm{AUROC}=\int_{0}^{1}\mathrm{TPR}(\mathrm{FPR})\,\diff\mathrm{FPR},
\end{equation}
where $\mathrm{TPR}$ and $\mathrm{FPR}$ are the true-positive and false-positive rates obtained by varying the score threshold for a chosen target (positive) class. An AUROC of $0.5$ corresponds to random guessing and $1.0$ to perfect separation. Equivalently,
\begin{equation}
\mathrm{AUROC}=\Pr\!\left(s(x^{+})>s(x^{-})\right)+\tfrac{1}{2}\Pr\!\left(s(x^{+})=s(x^{-})\right),
\end{equation}
i.e.\ the probability that a randomly chosen target event $x^{+}$ receives a higher score $s$ than a randomly chosen non-target event $x^{-}$ (with ties counted as $1/2$).
For both supervised and unsupervised methods we take the target class to be signal.

\begin{table}[htb]
\centering
\caption{Comparison of AUROC across algorithms. All algorithms are tested on the same independent test dataset (described in Sec.~\ref{sec:input}) composed of both signal and noise events. Errors denote one standard error estimated from 1000 stratified bootstrap resamples, resampling signal and noise events
independently with replacement.}
\label{tab:auroc_comparison}
\resizebox{\columnwidth}{!}{%
\begin{tabular}{l c c c c c}
\hline
\hline
\multirow{2}{*}{\textbf{Algorithm}} & \multirow{2}{*}{\textbf{Training data}} & \multicolumn{3}{c}{\textbf{Hit features}} & \multirow{2}{*}{\textbf{AUROC}} \\
\cline{3-5}
 & & \textbf{pos} & \textbf{time} & \textbf{charge} & \\
NHits & N/A & N/A & N/A & N/A & 0.8572 $\pm$ 0.0010 \\ \cline{3-6}
\multirow{4}{*}{\shortstack[l]{Supervised\\event-level\\classifier}} & \multirow{4}{*}{Mixed} & \checkmark & \checkmark & \checkmark & 0.8973 $\pm$ 0.0008 \\
 & & \checkmark & \checkmark & -- & 0.8928 $\pm$ 0.0008\\
 & & \checkmark & -- & \checkmark & 0.8878 $\pm$ 0.0009\\
 & & \checkmark & -- & -- & 0.8862 $\pm$ 0.0009\\ \cline{3-6}
\multirow{4}{*}{\shortstack[l]{Supervised\\hit-level\\classifier}} & \multirow{4}{*}{Mixed} & \checkmark & \checkmark & \checkmark & \textbf{0.9014} $\pm$ 0.0008\\
 & & \checkmark & \checkmark & -- & 0.9009 $\pm$ 0.0008\\
 & & \checkmark & -- & \checkmark & 0.8679 $\pm$ 0.0010\\
 & & \checkmark & -- & -- & 0.8678 $\pm$ 0.0010\\ \cline{3-6}
Autoencoder & Noise & \checkmark & \checkmark & -- & 0.7586 $\pm$ 0.0013 \\ \cline{3-6}
MPDR & Noise & \checkmark & \checkmark & -- & 0.8700 $\pm$ 0.0010 \\
\hline
\hline
\end{tabular}%
}
\end{table}

AUROC values are summarised in Table~\ref{tab:auroc_comparison} for all models
using all events in the testing samples, that is, covering the range of electron
kinetic energies from 0.5\,MeV to 7.0\,MeV. The supervised hit-level classifier gives the largest central AUROC value, $0.9014 \pm 0.0008$, followed by the supervised event-level classifier at $0.8973 \pm 0.0008$. Of the two anomaly-detection models, MPDR clearly outperforms the pure autoencoder, with AUROCs of $0.8700 \pm 0.0010$ and $0.7586 \pm 0.0013$, respectively. This puts the pure autoencoder behind NHits, which achieved $0.8572 \pm 0.0010$.

\begin{figure*}[htb]
    \centering
    \begin{subfigure}{\textwidth}
      \centering
      \includegraphics[height=0.12\textheight]{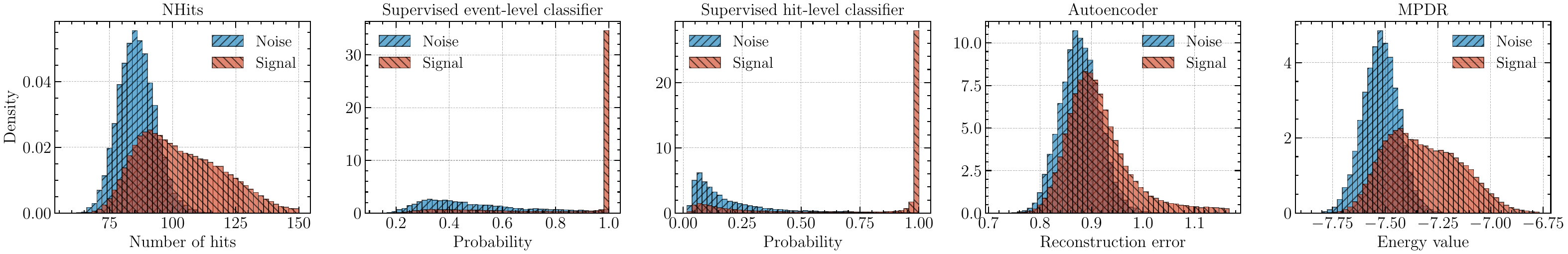}
      \caption{}
    \end{subfigure}

    \vspace{0.5em}

    \begin{subfigure}{\textwidth}
      \centering
      \includegraphics[height=0.12\textheight]{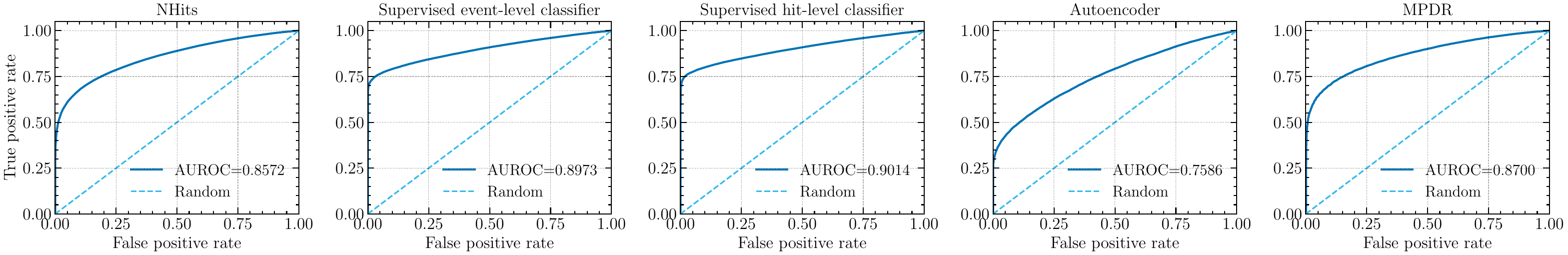}
      \caption{}
    \end{subfigure}

    \vspace{0.5em}

    \begin{subfigure}{\textwidth}
      \centering
      \includegraphics[height=0.12\textheight]{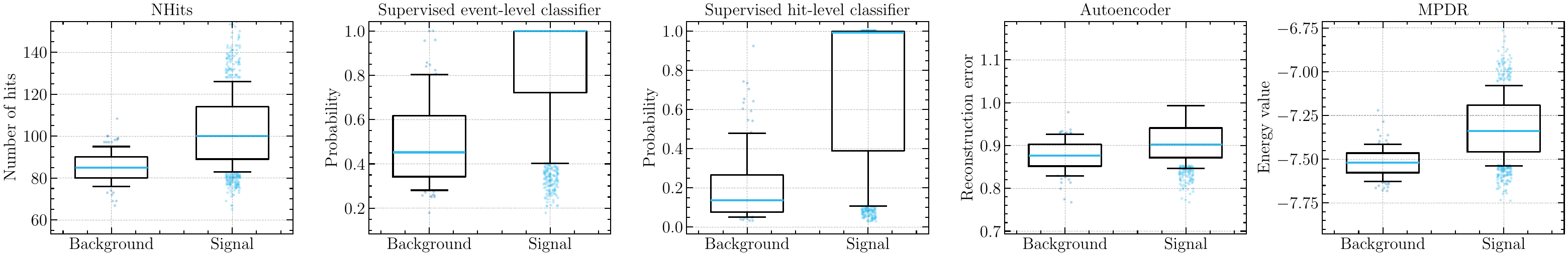}
      \caption{}
    \end{subfigure}
    
    \caption{\label{fig:distributions}
    Comparison of all five algorithms on noise and signal events. 
    (a): normalised histograms of score distributions for noise and signal. 
    (b): Receiver Operating Characteristic (ROC) curves summarising noise--signal separability.
    (c): boxplots of scores by class; in each boxplot, the central horizontal line denotes the median, the box spans the interquartile range, and the whiskers extend to the most extreme points within 1.5 times the interquartile range. Points beyond the whiskers are outliers; only a small fraction of outliers is shown to avoid overplotting.
    }
\end{figure*}

A visualisation of the separation power for each of the algorithms can be found
in Fig.~\ref{fig:distributions}: the top row of plots shows signal
and noise distributions of the relevant discriminator: the number of hits
for NHits, signal probabilities for the supervised classifiers, the
reconstruction error for the pure autoencoder, and the energy value for the
MPDR. Particularly striking are the significant spikes in the bin corresponding to
the highest probabilities for the supervised classifiers, and the more pronounced peak in the noise distribution for hit-level compared to event-level classification.

{The second row of plots in Fig.~\ref{fig:distributions} shows the ROC curves,
i.e.\ the true-positive rate (TPR) as a function of the false-positive rate
(FPR). Here, TPR corresponds to the signal efficiency, whereas FPR corresponds
to the fraction of noise events accepted by the trigger. Each point on a ROC curve corresponds to a
different threshold on the trigger score.}
The dependence on the FPR is particularly
informative as it is directly proportional to the false trigger rate, 
which has to be kept as low as possible to ensure realistic resource utilisation.
Both supervised classifiers keep a high TPR even at very low
FPR, with TPR around 0.75 at FPR $\approx$ 0.05, while NHits is close to 0.50 in
this low-FPR region, indicating much weaker discrimination. The MPDR achieves
values around 0.65 at FPR $= 0.05$.

Finally, the third row of Fig.~\ref{fig:distributions} shows boxplots for the
scores histogrammed in the top row, which further illustrates the strong
separation power of the supervised classifiers, in particular the hit-based
version. The impact of the MPDR in comparison to the pure autoencoder can also
clearly be seen in the stronger separation in signal and
noise in the corresponding plots.

Furthermore, we study alternative configurations of the two supervised
classifiers where only subsets of the available features are provided: in one
case only spatial and temporal information, in another case spatial information
and hit charge, and in a third case solely spatial information are used during
training and evaluation. Results can be found in
Table~\ref{tab:auroc_comparison} alongside the results for the standard configuration using all features shown above. For both supervised classifiers, hit times prove to be
more valuable than hit charges. Models without charge information perform similarly to the corresponding full-feature models: for the hit-level classifier, the AUROC changes from $0.9014 \pm 0.0008$ with all features to $0.9009 \pm 0.0008$ using position and time only. Models with spatial and charge features only on the other hand are more strongly impacted:
AUROCs are reduced to $0.8679 \pm 0.0010$ and $0.8878 \pm 0.0009$ for the hit-level and event-level-based
models, respectively, the event-level classifier thus outperforming the
hit-level equivalent in this case.  Models trained solely on spatial information show comparable performance to models incorporating both charge and spatial features, which further suggests the potential
to drop PMT charges as an input feature. Differences comparable to the quoted bootstrap uncertainty, such as the two hit-level configurations using all features and using position plus time only, should therefore not be over-interpreted.

\subsection{Signal efficiencies}

To compare trigger algorithms fairly, we evaluate all methods at threshold settings corresponding to an
operating point that ensures the false trigger rate remains below
$10~\mathrm{kHz}$. For each algorithm, this fixes a threshold, $\tau^\star$, on
its scalar trigger score $s$, determined from the independent noise-only
test sample. The corresponding signal efficiency is then the standard selection
efficiency evaluated at that threshold,
\begin{equation}
\varepsilon_{\mathrm{sig}}(\tau^\star)
= \frac{1}{N_{\mathrm{sig}}}\sum_{j=1}^{N_{\mathrm{sig}}}\mathbb{I}\!\left[s_j \ge \tau^\star\right],
\end{equation}
where $N_{\mathrm{sig}}$ is the number of signal events.

\begin{figure}[tb]
    \centering
    \includegraphics[height=.455\textwidth]{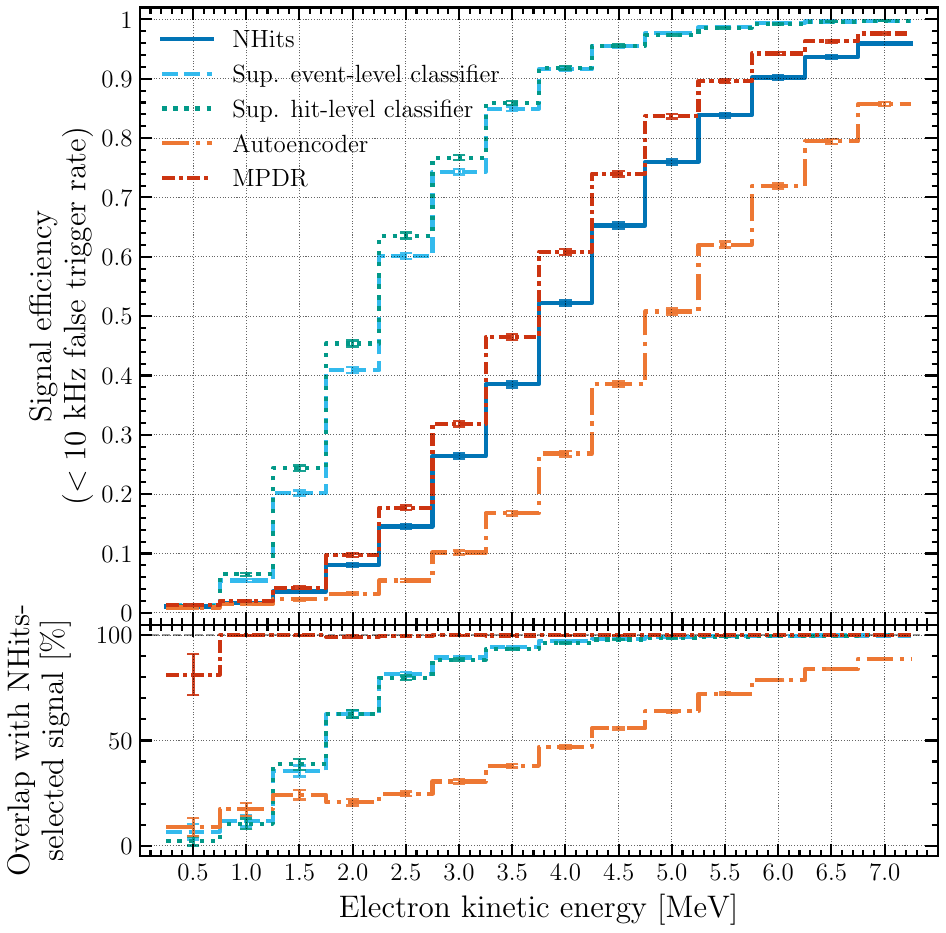}
    \caption{\label{fig:efficiencies}Signal tagging efficiencies for signals at different kinetic
    electron energies for all algorithms considered in this study. The lower panel shows the fraction of signal events selected by the NHits algorithm (Section~\ref{sec:evt_characteristics}) that were additionally also selected by the respective other algorithm, quantifying the complementarity. Error bars indicate binomial uncertainties.
    }
\end{figure}

Overall, the signal tagging efficiencies of all five algorithms at fixed false trigger
rate match what was observed with AUROC values, as can be seen in
Fig.~\ref{fig:efficiencies}. The supervised hit-level classifier exhibits best
results, with efficiencies of $(76.7\pm0.4)\%$ at 3\,MeV, $(91.9\pm0.3)\%$ at 4\,MeV, and
$(97.4\pm0.2)\%$ at 5\,MeV. The supervised event-level classifier follows at a similar
performance above 4\,MeV, however efficiencies are around 5-7\% lower in the
2-3\,MeV range compared to the hit-level equivalent. The MPDR outperforms NHits, with efficiencies of $(31.9\pm0.5)\%$ versus $(26.4\pm0.5)\%$ at
3\,MeV, $(60.8\pm0.5)\%$ versus $(52.2\pm0.5)\%$ at 4\,MeV, and $(83.7\pm0.4)\%$ versus $(76.0\pm0.4)\%$ at 5\,MeV. The pure autoencoder without
MPDR remains significantly less efficient than the NHits trigger, as can be seen
in the figure.

The lower panel of Fig.~\ref{fig:efficiencies} further shows the fraction of events identified as signal by both NHits and the given second algorithm as specified by colour code, relative to all events selected by NHits. This conditional overlap increases with electron energy for all methods, indicating that at higher energies the different trigger strategies increasingly identify the same, more prominent signals. At low energies, however, the overlap is substantially smaller for the supervised classifiers, which shows that their improved efficiency is not driven only by reproducing the NHits selection, but also by identifying additional low-energy signal events that NHits misses. The MPDR exhibits a larger overlap with NHits over much of the energy range than the supervised classifiers, indicating that a greater fraction of the signal events selected by NHits are also retained by MPDR while still improving on its efficiency. This also motivates further exploration of implementations that apply a combination of NHits and machine-learning-based algorithms.

\subsection{Efficiencies for neutron tagging on hydrogen}

The performance of these models is also evaluated in the context of neutron
tagging. In inverse beta decay interactions, antineutrinos produce neutrons that are later captured on free protons, resulting in the emission of a mono-energetic 2.2 MeV gamma ray. The time delay of the gamma relative to the prompt lepton signal is commonly exploited in neutron tagging algorithms. However, in this work, no temporal correlation with a preceding lepton signal is considered. The trained models described above are applied to these events,
and the signal identification efficiency is evaluated as before, requiring events with at
least one detected signal hit.  The supervised
hit-level classifier again demonstrates highest efficiencies, at 25.4\% for this
task. The event-level classifier follows at 21.8\%, but autoencoder and MPDR as
well as NHits remain below 4\%.  Results for all models are summarised in
Table~\ref{tab:gamma_eff}. 

\begin{table}[t]
\centering
\caption{Signal identification efficiencies for single, mono-energetic gammas at 2.2\,MeV,
when applying models trained on single electrons described above.}
\label{tab:gamma_eff}
\resizebox{\columnwidth}{!}{%
\begin{tabular}{l c c c c c}
\hline
\hline
\textbf{} &
\shortstack[c]{\vphantom{\textbf{Supervised}}\\ \textbf{NHits}\\ \vphantom{\textbf{classifier}}} &
\shortstack[c]{\textbf{Supervised}\\\textbf{event-level}\\\textbf{classifier}} &
\shortstack[c]{\textbf{Supervised}\\\textbf{hit-level}\\\textbf{classifier}} &
\shortstack[c]{\vphantom{\textbf{Supervised}}\\ \textbf{Autoencoder}\\ \vphantom{\textbf{classifier}}} &
\shortstack[c]{\vphantom{\textbf{Supervised}}\\ \textbf{MPDR}\\ \vphantom{\textbf{classifier}}} \\
\hline
\textbf{Efficiency} &
0.0417 &
0.2180 &
\textbf{0.2537} &
0.0204 &
0.0483 \\
\hline
\hline
\end{tabular}%
}
\end{table}

\subsection{Runtime estimates}

\begin{figure*}[htb]
    \centering
    \begin{subfigure}{\textwidth}
      \centering
      \includegraphics[height=0.16\textheight]{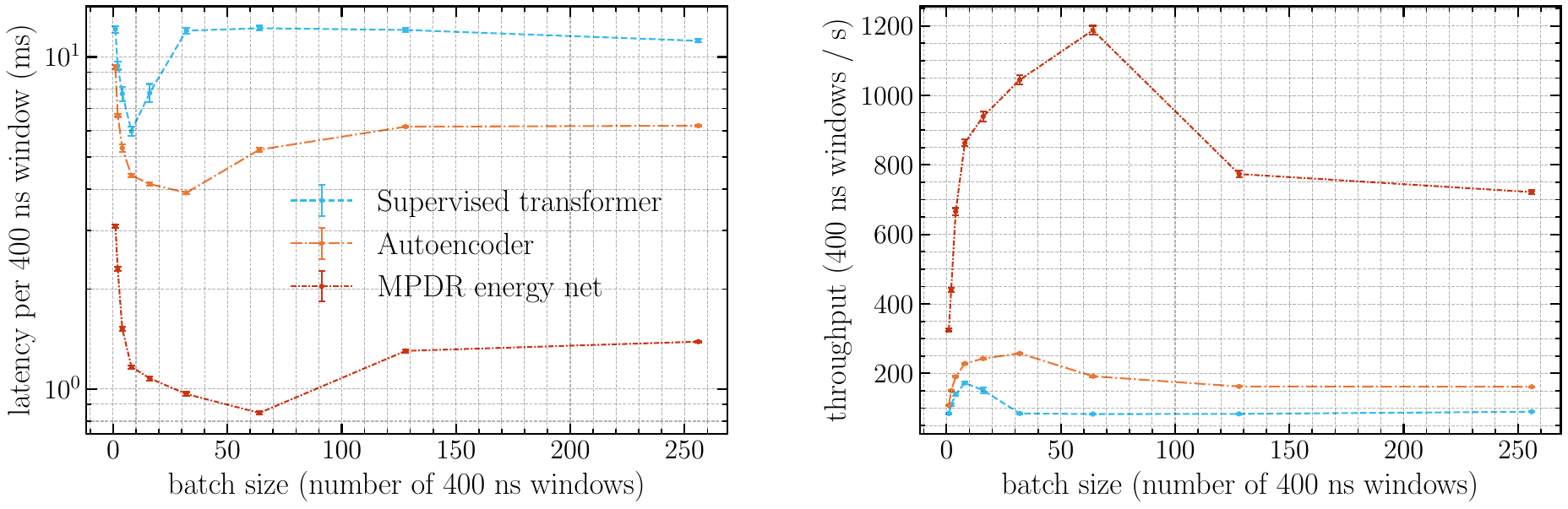}
      \caption{}
    \end{subfigure}

    \vspace{0.5em}

    \begin{subfigure}{\textwidth}
      \centering
      \includegraphics[height=0.16\textheight]{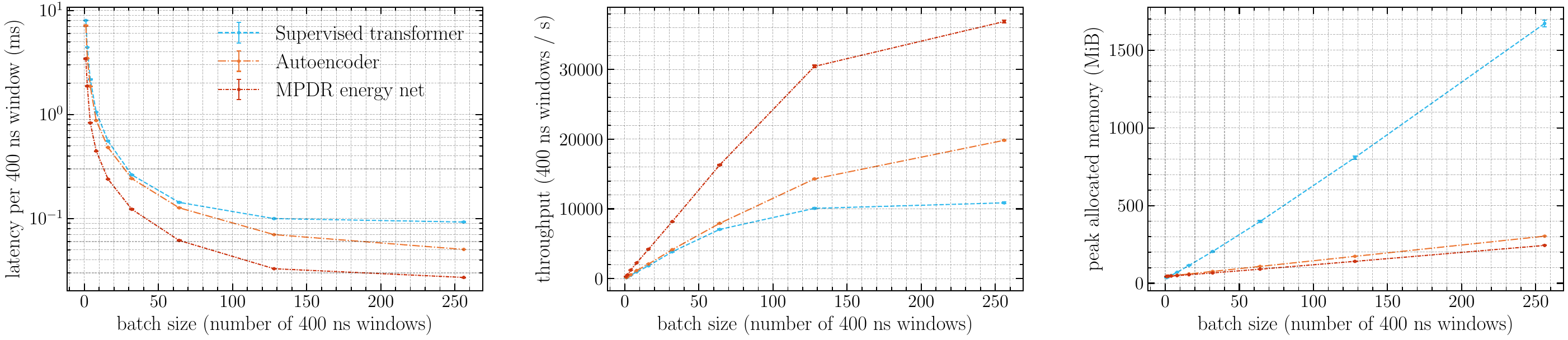}
      \caption{}
    \end{subfigure}
    \caption{\label{fig:batchsize}
    Inference performance as a function of batch size~$B$ (number of $400\,\mathrm{ns}$ windows per batch) for the Transformer (blue), autoencoder (orange), and energy network (red):
    (a) shows results obtained on CPU, and (b) shows results obtained on GPU.
    For each batch size, $50$ warm-up steps are performed, followed by $200$ timed steps.
    Left: latency per $400\,\mathrm{ns}$ window (per-item latency), shown on a logarithmic scale.
    Centre: throughput in windows per second.
    Right: peak allocated GPU memory during inference (MiB); GPU memory is only relevant for the GPU row.
    Error bars indicate the half-width of the two-sided $95\%$ confidence interval on the mean, estimated from the $200$ timed steps.
    }
\end{figure*}

We benchmark the inference performance of three neural-network architectures (the supervised transformer encoder used by both supervised classifiers, the autoencoder, and the MPDR energy network) as a function of
batch size~$B$, where each item in a batch corresponds to one fixed-duration input window of
$400\,\mathrm{ns}$.
For each configuration (model, device, batch size), we first perform 50 untimed preliminary inference calls, which are discarded from the benchmark. These calls ensure that one-off setup overheads, such as initial memory allocation and GPU kernel initialisation, do not bias the timing measurement. We then perform 200 timed inference steps.
Inference time is measured as wall-clock latency per step, including host-to-device transfer
(on GPU) and completion of the forward pass; on the GPU, explicit synchronisation
ensures that all queued device work is captured.
This end-to-end definition reflects the practical cost of applying the models in a data acquisition 
pipeline.
All benchmarks were performed on an AMD EPYC~7742 CPU (512\,GB host RAM) and an NVIDIA A100
80\,GB PCIe GPU.

For each configuration we record the per-step inference time and, on GPU, the peak allocated
memory during the step.
We report:
{(1)~the latency per window, obtained by dividing the per-step latency by the batch size~$B$;
(2)~the throughput, defined as the number of windows processed per second, i.e.\ $B$ divided by the per-step latency; and
(3)~the peak allocated GPU memory during inference, reported in MiB
($1\,\mathrm{MiB}=2^{20}$~bytes).}
Uncertainties correspond to the half-width of the two-sided 95\,\% confidence interval on the
mean, estimated from the $200$~timed steps using the Student's~$t$~distribution.

Figure~\ref{fig:batchsize} shows how batching affects inference cost.
For all three models, increasing~$B$ reduces the per-window latency and increases throughput, because fixed overheads are shared across more input windows. The size of this effect, however, depends strongly on both the model and the device.

On CPU (top row of Fig.~\ref{fig:batchsize}), the MPDR energy network gives the highest throughput over the full range of batch sizes. Its throughput increases rapidly at small batch size, reaches a maximum at an intermediate batch size, and then decreases for the largest batches shown. The autoencoder shows a weaker version of the same behaviour, with only modest gains from batching. The supervised transformer is substantially slower on CPU and benefits little from increasing the batch size. We therefore interpret the large-batch CPU region as a regime where increasing the batch size no longer improves the effective per-window processing rate for this hardware and implementation.

On GPU (bottom row of Fig.~\ref{fig:batchsize}), batching is much more effective. The per-window latency decreases sharply at small batch size for all models. At the largest batch sizes shown, the MPDR energy network reaches the highest throughput, of order $\sim3\times 10^4$ windows/s, followed by the autoencoder at roughly $2\times 10^4$ windows/s. The supervised transformer reaches a lower throughput, of order $10^4$ windows/s, and its curve flattens earlier than the other two. The rightmost panel shows that this comes together with a much larger GPU-memory footprint for the supervised transformer, while the autoencoder and MPDR energy network remain below a few hundred MiB over the range shown.

The larger runtime and memory cost of the supervised transformer is mainly caused by the way it uses hit-to-hit geometric information. For each event, the model does not only process the list of PMT hits individually; it also compares pairs of hits in order to include their relative positions and times in the attention calculation (Appendix~\ref{sec:rel_attention}). This provides useful information for recognising spatial and temporal patterns, but it also creates additional intermediate arrays whose size grows quickly with the number of hits. These arrays have to be stored and processed during inference, which increases both memory usage and runtime. In addition, the supervised transformer uses a larger input representation per hit than the autoencoder and MPDR energy network, since it includes the full PMT-level feature set. These effects explain why the supervised transformer is the most expensive model in Fig.~6, even though its number of trainable parameters is not larger.

On GPU, the measured per-window inference latencies are well below the millisecond scale and
reach the tens-of-microseconds range at large batch sizes, indicating substantial headroom for
near-real-time deployment.
Real-time performance would additionally depend on I/O, preprocessing, and scheduling, and is
not a target of this work.

\section{Conclusion}

We studied the performance of two custom network architectures for the task of
identifying low-energy signals in a large water Cherenkov detector such as
Hyper-Kamiokande. A supervised classifier was developed that overall showed best
performance when applying per-PMT predictions for signal and noise, using a
point-cloud representation of PMT positions, hit times and charges as inputs. A second
similar approach with event-level predictions performed similarly in the energy
range above 4\,MeV, but slightly worse at lower energies. Compared to a
hit-count based trigger as is custom for high-energy triggering in such
detectors, these models significantly improve the efficiency at identical false
trigger rates, thus allowing the detection boundary for low-energy neutrino
studies to be pushed to lower energies. The second model is an autoencoder-based anomaly detection approach
applied on a point-cloud representation of PMT positions and times. 
We compared a pure autoencoder with a more advanced version that additionally
applied MPDR, and found the latter outperforming the hit-count-based trigger as
well, at improvements in signal identification efficiencies of 5-8\% at identical false
trigger rate.  While the pure autoencoder was not found to be suitable for
this task, the model with MPDR demonstrates a first proof of concept of an
anomaly-detection-based low-energy trigger approach. Such a method is
particularly powerful as it is entirely agnostic to mis-modelling of signals,
provided the detector noise is understood and well-modelled. 

Moreover, variations of the supervised classifiers with reduced feature lists
were studied. Omitting hit charges in the features provided to the models did
not impact the results significantly, while omitting hit times showed a drastic
reduction in performance. Applying a trigger that operates on spatio-temporal
information only would allow to reduce the dependence on charge calibration of
the PMTs.

The identification of single gamma signals was studied as
well, where the models trained on single electrons were applied on events with
mono-energetic gammas at 2.2\,MeV, matching the energy in case of neutron
capture on hydrogen. The supervised hit-level classifier achieved a
gamma-identification efficiency of 25.4\% where a hit-count-based trigger only
achieves 4.2\% at identical false trigger rate, and the MPDR achieved an
efficiency of 4.8\% for this task. 

In addition to exhibiting higher efficiencies, trigger algorithms must
also satisfy stringent runtime performance requirements placed by the hardware
on-site, such that operation of the experiment in real-time is guaranteed.
Of the models compared here, the MPDR energy network showed highest throughput in runtime
studies on an NVIDIA A100 GPU, followed by the pure autoencoder.
This ordering reflects architectural differences: the MPDR energy network
performs a single encoder forward pass to produce a scalar score, whereas the
autoencoder requires both an encoder and a decoder pass for reconstruction.
The supervised transformer is the most expensive due to its custom relative-positional attention
(Appendix~\ref{sec:rel_attention}), which materialises explicit pairwise
coordinate differences in each layer.
To avoid unnecessary usage of computing resources, a hit-count-based pre-trigger
can be applied to ensure more computationally intensive algorithms are only
applied below a fixed hit-count threshold. Ultimately, feasibility to operate these models
as part of an online trigger suite depends on the available hardware.

The models presented here are implemented with runtime performance limitations
in mind, as operation as part of an online system is targeted. However,
potential offline reconstruction methods could be implemented based on similar
architectures. If runtime and false-trigger constraints are relaxed, even
higher signal identification efficiencies could be achieved, as required for a
detailed physics analysis. In addition, the gamma-identification performance discussed in the context of neutron tagging relies only on part of the available information. A dedicated model that incorporates both the lepton signal and the characteristic temporal delay of the gamma could likely outperform the results presented here.

Developments of machine-learning-based trigger algorithms such as presented in
this paper have great potential to help push the detection boundaries for
low-energy neutrino signals, as such algorithms excel at pattern recognition of
minimal signals on top of a noisy background. 
{Studies of solar and supernova neutrinos benefit greatly when data below 7\,MeV of kinetic electron energies can be acquired, and models presented here demonstrate the feasibility of efficient triggers down to 3\,MeV.}
{Moreover, a high neutron-tagging efficiency is crucial for Hyper-K, not only to allow improved studies of neutrinos in the low energy range, but also for measurements at higher energies. }

{The inclusion of machine-learning-based trigger algorithms in the Hyper-K DAQ framework will thus} help ensure that relevant
inputs can be provided to offline analyses, allowing dedicated measurements of extraterrestrial low-energy neutrinos at high precision.

\begin{acknowledgments}
This work was supported by the Swiss National Science Foundation under grants PCEFP2\textunderscore203261 and 200021L-231581; and by the UK Research and Innovation Grant MR/S032843/1.
Neural network training was supported through the Swiss AI Initiative via a grant from the Swiss National Supercomputing Centre (CSCS), project ID a149, on the Alps system. 

The authors also acknowledge fruitful discussions within the Hyper-Kamiokande DAQ working group and support during the preparation and review of this document, and would particularly like to thank Thomas Jones
for his interest in our work and for his thoughtful and constructive input. We would also like to thank the members of the Hyper-Kamiokande Editorial Committee for this manuscript for their valuable time and feedback.
\end{acknowledgments}

\section*{Author contributions}

K.L. and B.R. conceived the study. K.L. produced the simulated datasets, and implemented the baseline NHits trigger. S.A.-M. developed and validated the deep-learning-based trigger methods. K.L. and S.A.-M. performed the results analysis. D.S. and B.R. contributed through discussions and feedback throughout the project. All authors contributed to writing the manuscript.


\appendix

\section{Supervised classifier reproducibility}
\label{sec:appendix_transformer}

This appendix specifies the supervised transformer classifier implementation used in this work (Sec.~\ref{sec:supervised_classifier}).

\subsection{Geometry and normalisation}
Each hit provides Cartesian PMT coordinates $(x,y,z)$ on the detector surface, a hit time $t$ within the readout window, an integrated charge $q$, and a PMT-location label $\ell\in\{0,1,2\}$ indicating top endcap ($\ell=0$), barrel wall ($\ell=1$), or bottom endcap
($\ell=2$).

We convert positions to cylindrical coordinates,
\begin{equation}
\phi = \arctan2(y,x), \qquad
r = \sqrt{x^2+y^2},
\end{equation}
and use detector-aligned scalars for the embedding.
For wall PMTs we set $r=R$ (fixed barrel radius); for endcap PMTs we retain the physical $r$ on the endcap.

We normalise
\begin{equation}
\begin{aligned}
z_{\text{norm}} &= \frac{z-z_{\min}}{z_{\max}-z_{\min}}, \qquad
r_{\text{norm}} = \frac{r}{R}, \\
t_{\text{norm}} &= \frac{t-t_{\min}}{t_{\max}-t_{\min}},
\end{aligned}
\end{equation}
with $t_{\min}=0$~ns and $t_{\max}=400$~ns, and transform charge as
\begin{equation}
q_{\text{norm}} = \log(1+q).
\end{equation}

\subsection{Per-hit feature vector}
For each hit $i$, we build a feature vector $\mathbf{f}_i$ comprising:
\begin{itemize}
    \item \textbf{Base geometric features:}
    wall PMTs use $(\sin\phi,\,\cos\phi,\,z_{\text{norm}})$; endcap PMTs use $(r_{\text{norm}},\,\sin\phi,\,\cos\phi)$.
    \item \textbf{Fourier spatial encodings:}
    sinusoidal features with $K_z=K_r=4$ frequencies applied to $z_{\text{norm}}$ (wall) or $r_{\text{norm}}$ (cap), producing 8 features.
    \item \textbf{Fourier temporal encodings:}
    sinusoidal features with $K_t=6$ frequencies applied to $t_{\text{norm}}$, producing 12 features.
    \item \textbf{Charge:} the scalar $q_{\text{norm}}$.
\end{itemize}
Using all features yields $\mathbf{f}_i\in\mathbb{R}^{24}$.

\subsection{Location-aware input projection and special tokens}
To account for the distinct geometric roles of wall and endcap PMTs, we use separate learned linear input projections. For each hit $i$, the per-hit feature vector $f_i$ is mapped to a transformer token embedding $x_i \in \mathbb{R}^{d_{\rm model}}$ as
\begin{equation}
\begin{aligned}
\mathbf{x}_i &=
\begin{cases}
W_{\text{wall}}\mathbf{f}_i + \mathbf{b}_{\text{wall}}, & \ell_i=1,\\
W_{\text{cap}}\mathbf{f}_i + \mathbf{b}_{\text{cap}},  & \ell_i\in\{0,2\},
\end{cases} \\
&W_{\text{wall}},\,W_{\text{cap}} \in \mathbb{R}^{d_{\text{model}}\times d_{\text{feat}}},\\
&b_{\rm wall}, b_{\rm cap} \in \mathbb{R}^{d_{\rm model}}.
\end{aligned}
\end{equation}
where $W_{\rm wall}, W_{\rm cap}$ are learned projection matrices and $b_{\rm wall}, b_{\rm cap}$ are learned bias vectors. The projections are followed by token-type embeddings to distinguish top-cap, wall, and bottom-cap hits. Two special tokens are prepended: a learnable \texttt{[CLS]} token and a length token encoding the (unpadded) hit multiplicity through a learned linear map. Padding masks are applied in attention to support variable-length events.

\subsection{Transformer architecture}
We use a transformer encoder with $L=6$ layers, hidden size $d_{\text{model}}=64$, and $n_{\text{head}}=8$ attention heads (approximately $3\times10^5$ trainable parameters). Dropout with rate $p=0.1$ is applied to attention weights and feed-forward sublayers.

\subsection{Relative positional attention bias}
\label{sec:rel_attention}
Instead of absolute positional encodings, attention includes a learnable relative bias derived from pairwise hit differences.
For hits $i$ and $j$ we compute
\begin{equation}
\Delta r_{ij}=r_i-r_j,\qquad
\Delta z_{ij}=z_i-z_j,\qquad
\Delta t_{ij}=|t_i-t_j|,
\end{equation}
and an azimuthal difference wrapped to the principal interval using a two-component arctangent,
\begin{equation}
\Delta \phi_{ij}=\mathrm{wrap}(\phi_i-\phi_j)
=\textrm{arctan2}\!\big(\sin(\phi_i-\phi_j),\cos(\phi_i-\phi_j)\big).
\end{equation}
These features are passed through a Multi Layer Perception (MLP) to produce per-head biases,
\begin{equation}
\mathbf{b}_{ij}=\mathrm{MLP}_{\text{bias}}\!\left([\Delta r_{ij},\Delta\phi_{ij},\Delta z_{ij},\Delta t_{ij}]\right)\in\mathbb{R}^{n_{\text{head}}}.
\end{equation}
The attention logits become
\begin{equation}
\mathrm{Attention}(Q,K,V)=\mathrm{softmax}\!\left(\frac{QK^{T}}{\sqrt{d_k}}+\alpha\,\mathbf{B}\right)V,
\end{equation}
where $\mathbf{B}\in\mathbb{R}^{n_{\text{head}}\times N\times N}$ stacks the $\mathbf{b}_{ij}$ and $\alpha$ is a learned scale.

\subsection{Training and optimisation}
The model is trained for 50 epochs with batch size 256 using AdamW~\cite{loshchilov2017decoupled}
($\beta_1=0.9$, $\beta_2=0.95$) and learning rate $10^{-4}$. We use a cosine annealing schedule~\cite{loshchilov2016sgdr} with a 5-epoch linear warm-up followed by cosine decay over the remaining epochs. Mixed-precision (bfloat16) training is used on two NVIDIA GH200 GPUs.

For event-level supervision, binary cross-entropy is applied to the event label. For hit-level supervision, binary cross-entropy is applied to the unmasked per-hit logits, with labels indicating whether each PMT hit (token) originates from Cherenkov signal or detector noise.

\subsection{Feature configurations}
To quantify the importance of different input modalities, we train separate models with:
\begin{itemize}
    \item \textbf{All features:} spatial + time + charge (${d_{\text{feat}}=24}$),
    \item \textbf{Position and charge:} spatial + charge (${d_{\text{feat}}=12}$),
    \item \textbf{Position and time:} spatial + time (${d_{\text{feat}}=23}$),
    \item \textbf{Positions only:} spatial only (${d_{\text{feat}}=11}$).
\end{itemize}

\section{MPDR reproducibility}
\label{sec:appendix_anomaly}

This appendix specifies the MPDR implementation used in this work. In the reported configuration we use feature mode without charge (position and time only), so an event is a variable-length point cloud
$X\in\mathbb{R}^{N\times 4}$ with per-hit features $(u_x,u_y,u_z,t)$, where $\mathbf{u}\in\mathbb{S}^2$ is obtained by rescaling detector coordinates, then projecting to unit norm. For a detector centred at $z=0$, this is equivalently written as $(x/R,\,y/R,\,z/H)$ with $H=(z_{\max}-z_{\min})/2$, and
$t\in[-1,1]$ is the normalised hit time (Sec.~\ref{sec:mpdr}). The decoder outputs a fixed number $M$ of candidate hits
($M=192$ in our implementation), each with an existence probability; variable cardinality is handled by these probabilities and masking.

\subsection{Network architectures}
\label{sec:appendix_arch}

We use transformer blocks consisting of multi-head self-attention followed by a position-wise feed-forward network (FFN). Unless stated otherwise, the model dimension is 128, the FFN hidden dimension is 512, and dropout is set to 0.0.

\subsubsection{Transformer encoder (autoencoder)}
The encoder $\mathcal{E}$ maps a variable-length set of hits to a latent vector.
\begin{itemize}
    \item \textbf{Embedding:} linear map $4\rightarrow 128$ applied per hit.
    \item \textbf{Backbone:} 4 transformer layers with 8 attention heads; padding-masked attention for variable-length inputs.
    \item \textbf{Pooling:} a learnable CLS token is prepended; the final CLS embedding is taken as $\mathbf{z}\in\mathbb{R}^{128}$.
    \item \textbf{Latent normalisation:} $\mathbf{z}\leftarrow \mathbf{z}/\|\mathbf{z}\|$ (unit hypersphere; unit-norm latent constraint).
\end{itemize}

\subsubsection{Transformer decoder (autoencoder)}
The decoder $\mathcal{D}$ maps $\mathbf{z}$ to $M$ candidate hits and their existence probabilities.
\begin{itemize}
    \item \textbf{Queries:} $M=192$ learned query embeddings $\{\mathbf{q}_j\}_{j=1}^M\subset\mathbb{R}^{128}$.
    \item \textbf{Latent conditioning:} 4 learned memory tokens produced from $\mathbf{z}$ by linear maps.
    \item \textbf{Backbone:} 3 transformer decoder layers with (i) self-attention over queries and (ii) cross-attention to memory tokens (8 heads).
    \item \textbf{Heads (per query):}
    \begin{itemize}
        \item spatial direction from a raw 3-vector $\mathbf{y}^{\text{raw}}_j$ via $\mathbf{u}_j=\mathbf{y}^{\text{raw}}_j/\|\mathbf{y}^{\text{raw}}_j\|$,
        \item time $\tau_j=\tanh(\cdot)\in[-1,1]$,
        \item existence probability $p_j=\sigma(\cdot)\in(0,1)$.
    \end{itemize}
\end{itemize}

\subsubsection{Energy network (MPDR-S)}

For MPDR-S, the energy function $E_\theta(X)$ uses the same transformer-encoder layout as $\mathcal{E}$ (4 layers, 8 heads, 128D), but
with LeakyReLU (slope 0.2) in the FFN, following the recommendation of Ref.~\cite{yoon2023energybasedmodelsanomalydetection}; we observed similar behaviour in our experiments.
In place of CLS pooling, the energy network uses top-$k$ log-mean-exp pooling ($k=10$, $\tau=0.1$) over per-token scalar energies.
The final linear head $128\rightarrow 1$ is regularised with spectral normalisation and produces a scalar energy.

\subsection{Reconstruction loss: multi-scale density-aware Chamfer distance}
\label{sec:appendix_dcd}

Our loss is a custom implementation of the density-aware Chamfer distance~\cite{wu2021densityawarechamferdistancecomprehensive}, tailored for our specific task.
Let the input be $X=\{(\mathbf{u}_i,t_i)\}_{i=1}^N$ and the decoder output be $(Y,\mathbf{p})$ with
$Y=\{(\mathbf{v}_j,\tau_j)\}_{j=1}^M$ and $\mathbf{p}=\{p_j\}_{j=1}^M$.
We define squared distances between spatial directions
\begin{equation}
d_{ij}^2=\|\mathbf{u}_i-\mathbf{v}_j\|^2,
\end{equation}
and a weighted multi-scale kernel (three Gaussian scales)
\begin{equation}
\begin{aligned}
k_{ij} &= \sum_{\ell=1}^{3}\omega_\ell \exp(-\alpha_\ell d_{ij}^2),\\
\boldsymbol{\alpha} &= (5,30,120), \qquad
\boldsymbol{\omega} = (0.6,0.3,0.1).
\end{aligned}
\end{equation}

\subsubsection{Soft correspondences and density factors}
Hard nearest-neighbour assignments are replaced by soft correspondences
\begin{align}
w_{ij} &= \frac{\exp(-\beta d_{ij}^2)}{\sum_{j'}\exp(-\beta d_{ij'}^2)}, &
\tilde w_{ji} &= \frac{\exp(-\beta d_{ij}^2)}{\sum_{i'}\exp(-\beta d_{i'j}^2)},
\end{align}
with a batch-adapted temperature $\beta$ clamped to $[5,200]$.
To reduce sensitivity to local point density we use
\begin{equation}
c_j=\sum_i w_{ij},\qquad
\tilde c_i=\sum_j p_j\tilde w_{ji},\qquad
\lambda=1.
\end{equation}

\subsubsection{Loss terms}
\textbf{Recall (input coverage):}
\begin{equation}
\mathcal{L}_{\text{recall}}=
\frac{1}{N}\sum_{i=1}^{N}\left[1-\sum_{j=1}^{M} w_{ij}\,k_{ij}\,p_j\,c_j^{-\lambda}\right].
\end{equation}

\textbf{Precision (reconstruction quality):}
\begin{equation}
\mathcal{L}_{\text{precision}}=
\frac{1}{\sum_j p_j}\sum_{j=1}^{M} p_j\left[1-\sum_{i=1}^{N} \tilde w_{ji}\,k_{ij}\,\tilde c_i^{-\lambda}\right].
\end{equation}

\textbf{Cardinality:}
\begin{equation}
\mathcal{L}_{\text{count}}=\text{SmoothL1}\!\left(\sum_{j=1}^{M}p_j-N,\ \delta=5\right).
\end{equation}

\textbf{Time matching:} using kernel-weighted correspondences, we match times in both directions. For input$\rightarrow$reconstruction,
\begin{align}
t_i^{\text{matched}} &= \sum_j
\bar w_{ij}\,\tau_j,\qquad
\bar w_{ij} = \frac{w_{ij}k_{ij}}{\sum_{j'} w_{ij'}k_{ij'}},
                   \\
\mathcal{L}^{(a)}_{\text{time}} &= \frac{1}{N}\sum_i |t_i-t_i^{\text{matched}}|,
\end{align}
and analogously, for reconstruction$\rightarrow$input,
\begin{align}
\tau_j^{\text{matched}} &= \sum_i
\bar{\tilde w}_{ji}\,t_i,\qquad
\bar{\tilde w}_{ji} = \frac{\tilde w_{ji}k_{ij}}{\sum_{i'} \tilde w_{ji'}k_{i'j}},
\\
\mathcal{L}^{(b)}_{\text{time}} &= \frac{1}{\sum_j p_j}\sum_j p_j\,|\tau_j-\tau_j^{\text{matched}}|,
\end{align}
and we set
$\mathcal{L}_{\text{time}}=\tfrac{1}{2}(\mathcal{L}^{(a)}_{\text{time}}+\mathcal{L}^{(b)}_{\text{time}})$.

\vspace{0.4cm}
\textbf{Repulsion:} to discourage collapse of the decoded candidates, we use a nearest-neighbour penalty
\begin{equation}
\begin{aligned}
\delta_{j,\mathrm{nn}}^2 &= \min_{j'\neq j}\|\mathbf{v}_j-\mathbf{v}_{j'}\|^2,\\
\mathcal{L}_{\text{repulsion}} &=
\frac{1}{\sum_j p_j}\sum_{j=1}^{M} p_j\exp\!\left(-\gamma\,\delta_{j,\mathrm{nn}}^2\right),\\
&\qquad \gamma=400.
\end{aligned}
\end{equation}

\textbf{Total:}
\begin{equation}
\begin{split}
\mathcal{L}_{\text{AE}} ={} & \mathcal{L}_{\text{recall}} + \mathcal{L}_{\text{precision}} + 0.1\,\mathcal{L}_{\text{count}} \\
& + 0.5\,\mathcal{L}_{\text{time}} + 0.02\,\mathcal{L}_{\text{repulsion}}.
\end{split}
\end{equation}

\subsection{MPDR training: negative sample generation}
\label{sec:appendix_mpdr}

\subsubsection{Cardinality for negatives}

In the reported training configuration, each negative sample is assigned the same cardinality as its corresponding positive event ($K_{\text{neg}}=K_{\text{pos}}$), and the top-$K_{\text{neg}}$ decoder candidates ranked by existence probability $p_j$
are retained (the remainder are masked as padding).

\subsubsection{Latent perturbation and recovery}

Given a noise event $X_{\text{pos}}$, define $\mathbf{z}_{0}=\mathcal{E}(X_{\text{pos}})$ and form the latent reference
\begin{equation}
\begin{aligned}
\mathbf{z}_{\text{ref}}
&=
\frac{\mathbf{z}_{0}+\sigma_{\text{proj}}\boldsymbol{\epsilon}}
{\|\mathbf{z}_{0}+\sigma_{\text{proj}}\boldsymbol{\epsilon}\|},
\qquad \boldsymbol{\epsilon}\sim\mathcal{N}(0,I), \\
&\qquad \sigma_{\text{proj}}\sim\text{Uniform}(0.06,0.15).
\end{aligned}
\end{equation}
The decoded event $\mathcal{D}(\mathbf{z}_{\text{ref}})$ provides the initial synthetic sample, and the same $\mathbf{z}_{\text{ref}}$ is used in the recovery penalty.

\textbf{Latent initialisation (2 steps):}
\begin{equation}
\mathbf{z}^{(m+1)}=
\Pi_{\mathbb{S}^{127}}\!\left[\mathbf{z}^{(m)}-\eta_z\nabla_{\mathbf{z}}J_z(\mathbf{z}^{(m)})+\sigma_z\boldsymbol{\epsilon}_m\right],
\end{equation}
where $\Pi_{\mathbb{S}^{127}}$ normalises to unit length, $\mathbf{z}^{(0)}=\mathbf{z}_{\text{ref}}$, $\eta_z=0.01$, $\sigma_z=0.01$, and
\begin{equation}
J_z(\mathbf{z})=
\frac{1}{0.8}E_\theta(\mathcal{D}(\mathbf{z}))+\frac{0.1}{2\,\sigma_{\text{proj}}^2}\arccos^2(\mathbf{z}\cdot\mathbf{z}_{\text{ref}}).
\end{equation}

\textbf{Visible-space recovery (10 steps):}
letting $X_{\text{init}}$ denote the decoded sample after the latent initialisation stage (or $\mathcal{D}(\mathbf{z}_{\text{ref}})$ if this stage is skipped), we update
\begin{equation}
X_{\text{raw}}^{(m+1)}=
\Pi_{\text{model}}\!\left[X_{\text{raw}}^{(m)}-\eta_x\nabla_X J_x(X_{\text{raw}}^{(m)})+\sigma_x\boldsymbol{\epsilon}_m\right],
\end{equation}
with $\sigma_x=0.02$ and $\eta_x=80.0\,\sigma_x^2/2=0.016$. Here $\Pi_{\text{model}}$ renormalises the first three coordinates to unit norm and reflectively wraps the auxiliary features into $[-1,1]$. The objective is
\begin{equation}
\begin{aligned}
J_x(X_{\text{raw}})= {} &
\frac{1}{0.8}E_\theta(\Pi_{\text{model}}(X_{\text{raw}})) \\
& +
\frac{0.1}{2\,\sigma_{\text{proj}}^2}
\arccos^2\!\left(
\mathcal{E}(\Pi_{\text{model}}(X_{\text{raw}}))\cdot\mathbf{z}_{\text{ref}}
\right).
\end{aligned}
\end{equation}

\subsection{Optimisation and preprocessing}
\label{sec:appendix_preproc}

\subsubsection{Optimisation}
\textbf{Phase I (autoencoder):} AdamW ($\beta_1=0.9$, $\beta_2=0.999$, $\epsilon=10^{-8}$),
learning rate $10^{-4}$ with cosine annealing and 5-epoch warm-up, weight decay $10^{-4}$, batch size 512,
gradient clipping 1.0, 200 epochs (80/20 train/validation split).

\textbf{Phase II (energy):} AdamW with the same $\beta$ values, learning rate $5\times 10^{-5}$ with cosine annealing and 5-epoch warm-up,
weight decay 0.0, batch size 512, gradient clipping 1.0, 50 epochs. The contrastive divergence loss uses a softplus temperature $\tau=10$
and regularisation coefficients $\gamma_{\mathrm{vx}}=10^{-4}$ (squared-energy penalty), $\lambda_{\text{centre}}=10^{-4}$ (centring), and $\lambda_{\text{L2}}=10^{-4}$ (weight norm).

\subsubsection{Spatial and temporal normalisation}
For a hit at Cartesian position $(x,y,z)$ on the detector surface, we first rescale
\begin{equation}
\tilde x=\frac{x}{R},\qquad \tilde y=\frac{y}{R},\qquad \tilde z=2\cdot\frac{z-z_{\min}}{z_{\max}-z_{\min}}-1.
\end{equation}
and then define
\begin{equation}
\mathbf{u}=\frac{(\tilde x,\tilde y,\tilde z)}{\sqrt{\tilde x^2+\tilde y^2+\tilde z^2}}.
\end{equation}

Hit times $t\in[0,400]$~ns are normalised to $[-1,1]$ via
\begin{equation}
t_{\text{norm}} = 2\cdot\frac{t}{400}-1.
\end{equation}

\section{Augmentations}
\label{sec:appendix_augmentations}
The same geometric augmentations are applied per event during supervised training and autoencoder pre-training:
random rotation about the detector axis (azimuthal angle uniform in $[0,360^\circ)$) and independent reflections of the Cartesian detector coordinates along each axis, with 50\% probability per axis, before the final feature construction.

\bibliographystyle{apsrev4-1}
\bibliography{references}

\end{document}